\documentclass[reprint, superscriptaddress, nofootinbib,aps,prl]{revtex4-2}
\usepackage{hyperref}
\usepackage{amsmath}
\usepackage{amssymb}
\usepackage{graphicx}
\usepackage{dcolumn}
\usepackage{bm}
\usepackage[inline]{enumitem}
\usepackage{xcolor}
\usepackage{dsfont}
\usepackage{verbatim}
\usepackage{mathtools, xcolor,ytableau, amsfonts,tikz}

\newcommand{\mub}{\bar{\mu}}
\newcommand{\pd}{\partial}

\begin{document}

\preprint{APS/123-QED}

\title{Chiral Modes of Giant Superfluid Vortices}

\author{Gabriel Cuomo}
\email{gc6696@princeton.edu}
\affiliation{Center for Cosmology and Particle Physics, Department of Physics, New York University, New York, NY 10003, USA}
\affiliation{Department of Physics, Princeton University, Princeton, New Jersey 08544, USA}
\affiliation{Simons Center for Geometry and Physics, Stony Brook University, Stony Brook, New York 11794, USA}
\affiliation{C. N. Yang Institute for Theoretical Physics, Stony Brook University, Stony Brook, New York 11794, USA}

\author{Zohar Komargodski}
\email{zkomargodski@scgp.stonybrook.edu}
\affiliation{Simons Center for Geometry and Physics, Stony Brook University, Stony Brook, New York 11794, USA}

\author{Siwei Zhong}
 \email{siwei.zhong@stonybrook.edu}
\affiliation{Simons Center for Geometry and Physics, Stony Brook University, Stony Brook, New York 11794, USA}
\affiliation{C. N. Yang Institute for Theoretical Physics, Stony Brook University, Stony Brook, New York 11794, USA}

\date{\today}

\begin{abstract}
We discuss the rapidly rotating states of a superfluid. We concentrate on the giant-vortex (GV) state, which is a coherent rotating solution with a macroscopic hole at the center. We show that, for any trap, the fluctuations obey an approximately chiral dispersion relation, describing arbitrary shape deformations moving with the speed of the ambient superfluid. This dispersion relation is a consequence of a peculiar infinite symmetry group that emerges at large angular velocity and implies an infinite ground-state degeneracy. The degeneracy is lifted by small corrections which we determine for general equations of state and trapping potentials.
\end{abstract}

\maketitle

\section{Introduction}

Many systems with $U(1)$ symmetry break the symmetry spontaneously at low temperatures and at finite density. In particular, this was observed in ${}^4\text{He}$ and in trapped alkali-metal gases, which are in a superfluid phase at low temperatures. See~\cite{fetter2001vortices, pethick2008bose, Schmitt:2014eka}, among many others, for reviews and references. Non-rotating
superfluids contained in a trap have low-energy excitations with a linear dispersion relation, where $c_\text{s}$ is the sound speed.

New physics emerges when the trap rotates and superfluids are stirred~\cite{2008LaPhy..18....1F}. In two spatial dimensions, as the frequency of the trap increases, vortices appear \cite{madison2000vortex} and form an Abrikosov lattice~\cite{doi:10.1126/science.1060182}. In such a lattice, superfluid excitations (Tkachenko modes~\cite{tkachenko1966vortex,tkachenko1966stability,tkachenko1969elasticity,sonin2014tkachenko}) have the dispersion relation $\omega \sim \vec k^2$ (see \cite{Moroz:2018noc,Du:2022xys} for a modern approach). 

\begin{figure}[!h]
\includegraphics[width=0.23\textwidth ]{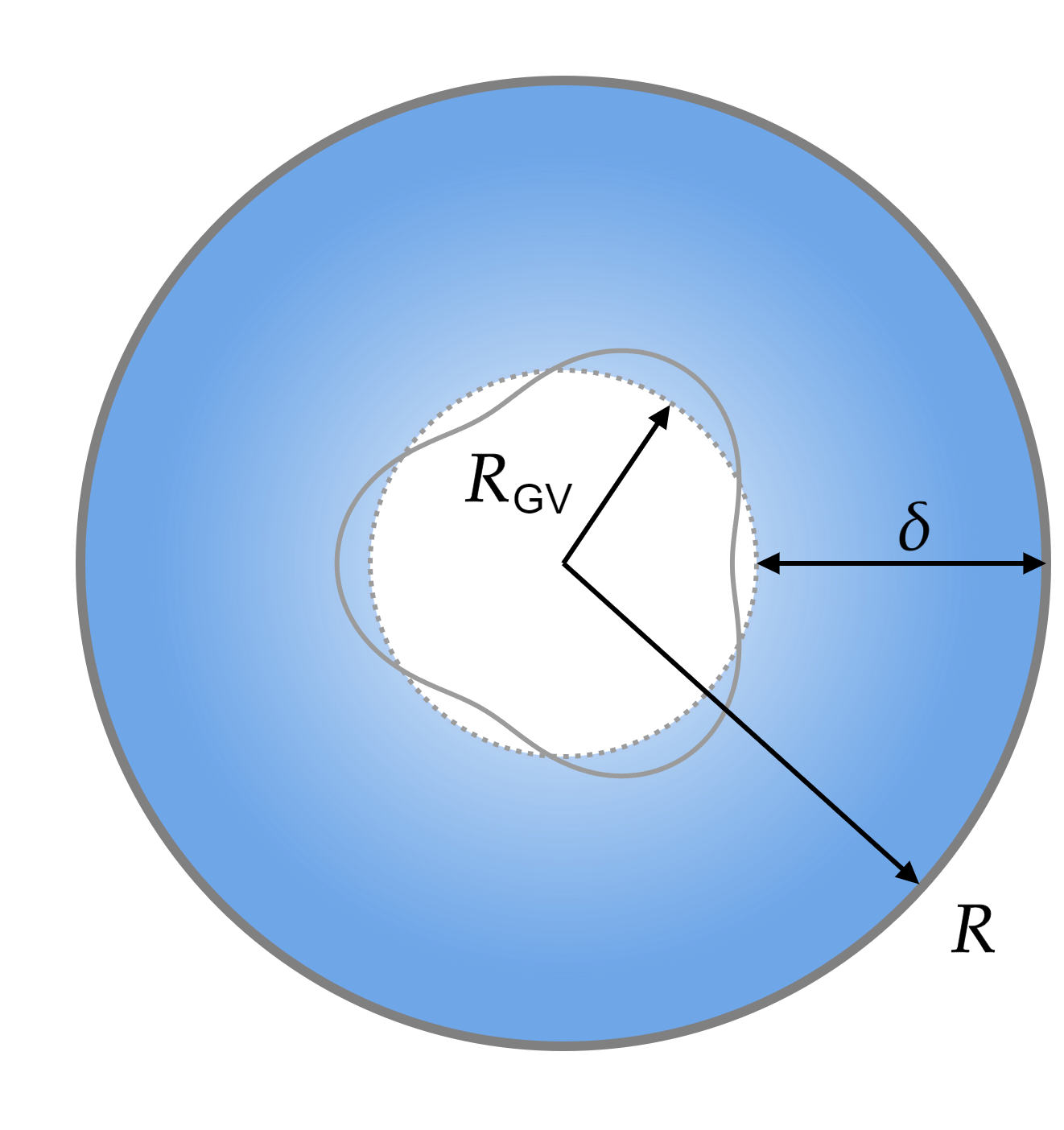}
\includegraphics[width=0.23\textwidth ]{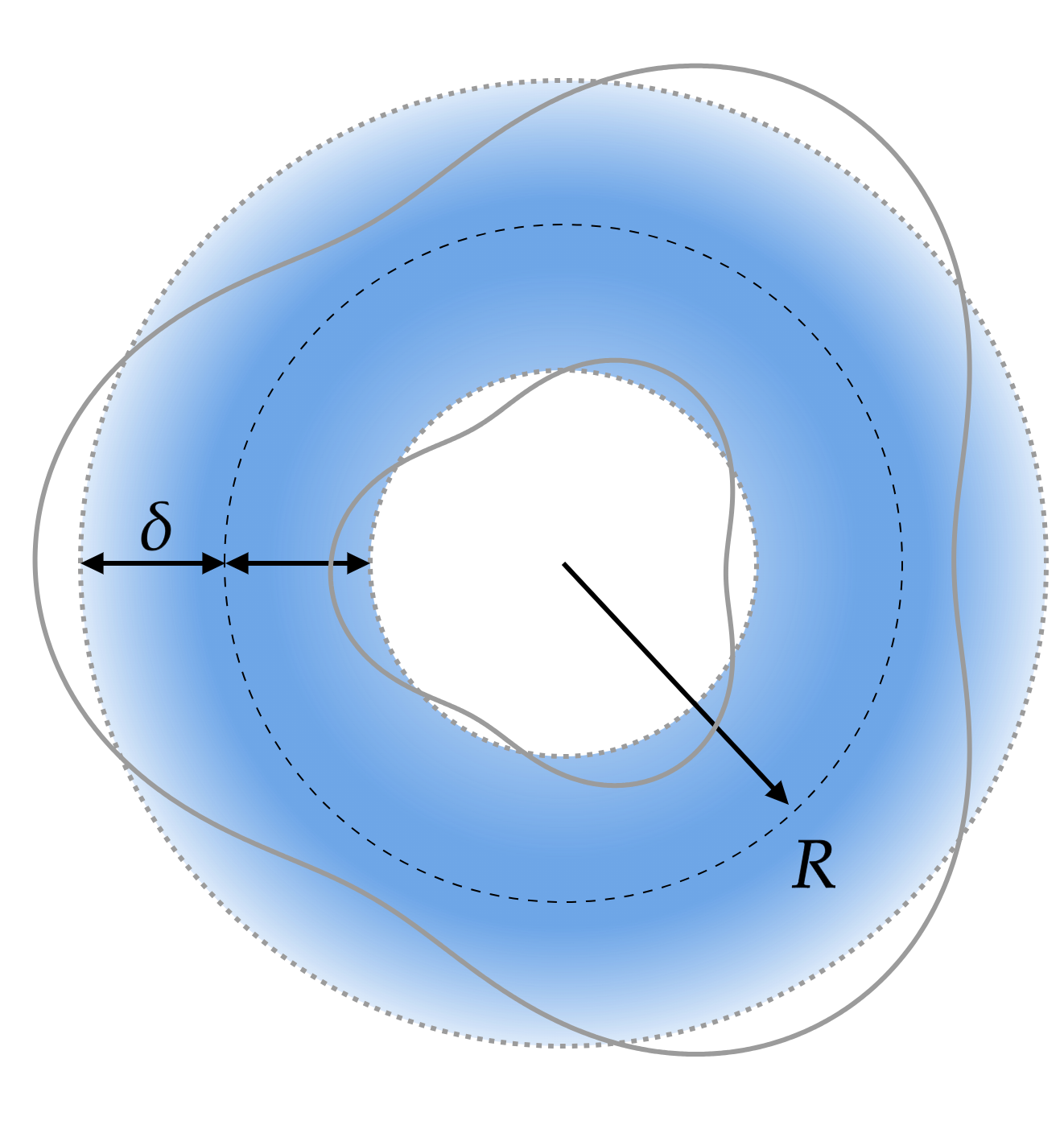}
\caption{The GV of width $\delta$ and radius $R_{\text{GV}}$ in a hard cylindrical trap of radius 
$R$ (left) and in a smooth trap (right). Density fluctuations of the GV deform the shape of the edges and move together with the trap. }\label{plot_GV parameter and chiral fluctuation}
\end{figure}

Above a certain trap frequency superfluids are expected to enter new phases. Here we concentrate on the giant vortex (GV)~\cite{fischer2003vortex,2004PhRvA..69c3608A,fetter2005rapid,correggi2013vortex, Cuomo:2022kio}. This configuration exists for superfluids rotating with supersonic velocity in anharmonic traps \cite{guo2020supersonic}, for which the centrifugal force dynamically induces a large hole. The fluid is localized over an annulus with no vorticity in the bulk (see Fig.~\ref{plot_GV parameter and chiral fluctuation}). This configuration is expected to be stable at large angular velocity~\cite{fetter2005rapid}. As a side remark, the term ``giant vortex" is also used in different contexts \cite{gauthier2019giant}, particularly for certain solitons in superconductors \cite{tanaka2002electronic, dao2011giant,  palonen2013giant}. These solitons have similarities to the superfluid GVs that we consider in this work, but are physically distinct since a dynamical gauge field is present in the former.

We study small density fluctuations of the (narrow) giant vortex,  as shown in Fig.~\ref{plot_GV parameter and chiral fluctuation}.  We find that perturbations are co-moving with the GV in the limit of large rotation speed $\Omega$:
\begin{equation}\label{intrdis} 
\omega \simeq \Omega n~,
\end{equation}
where $n\in \mathbb{Z}$ is the angular momentum of the density perturbation and $\omega$ is the corresponding frequency. The dispersion relation \eqref{intrdis} describes chiral waves, i.e., unidirectional perturbations of the fluid. Note that the GV breaks time-reversal symmetry, and the chiral modes bear similarity to those of various systems, such as the relativistic chiral boson \cite{PhysRevLett.59.1873,elitzur1989remarks}. The fact that $n$ appears without absolute value leads to a peculiar infinite degeneracy of the ground state at fixed angular momentum. Indeed, Fock space states with $\sum n_i=0$ have the same quantum numbers and energy as the unperturbed GV. 

As we make $\Omega$ larger at a fixed particle number,  the radius $R$ of the GV increases due to the centrifugal force, while the thickness $\delta$ becomes smaller. The lowest-lying excitations are approximately constant over the annulus radial direction. Therefore, the effective theory of the fluctuations leading to~\eqref{intrdis} lives in one spatial dimension and describes the physics to leading order in $\delta/R$. The infinite degeneracy originates from peculiar symmetries of this one-dimensional effective field theory (EFT). At leading order in $\delta/R$ the EFT is invariant under the warped conformal group~\cite{Jensen:2017tnb}, as well as under a peculiar fractonic symmetry, reminiscent of \cite{2002PhRvB..66e4526P, Burnell:2021reh}.

A useful way to think about~\eqref{intrdis} is that, in the rotating frame, the superfluid has an approximately vanishing speed of sound.
In reality, the speed of sound scales as $\delta \Omega\ll R\Omega$. This leads to $O(\delta/R)$ small corrections to the dispersion relation, by which the ground state degeneracy is lifted. We explicitly compute such corrections to~\eqref{intrdis}. We find that they depend on certain details of the trap, in particular, its steepness. We consider both smooth traps (generic axisymmetric continuous potentials $V(r)$) and the hard trap (a hard wall that confines the superfluid in the region $r<R$).

The rest of the text is organized as follows. First, we review the effective field theory approach to superfluids. We then discuss the properties of the giant-vortex solution. The fluctuation spectrum of the giant vortex is investigated in the following sections. In particular, we study the spectrum of fluctuations for the Gross-Pitaevskii model in a hard trap, models with a generic equation of state $P(X)$ in a hard trap, and generic models in smooth traps. We finally discuss the one-dimensional effective theory of the chiral modes.

\section{Superfluid Effective Theory}

We consider Galilean invariant superfluids in two-dimensional space. At low energies we may neglect derivatives of the density field, and the low-energy physics of these systems is described in terms of the Nambu-Goldstone boson $\phi \in \mathbb{S}^1$, which is the phase of the $U(1)$ order parameter. To the lowest order in derivatives, the effective action is written in terms of $\phi$ via the combination~\cite{Son:2002zn,Son:2005rv} (we set $\hbar=1$ throughout) 
\begin{equation}
\label{eq_functional argument}
    \begin{aligned}
    X\equiv \partial_t \phi-V(x)-\frac{(\nabla \phi)^2}{2m}~,
    \end{aligned}
\end{equation}
where $m$ is the mass of the microscopic constituents and $V(x)$ is the potential of the trap. $V(x)$ can be viewed as a background gauge field coupled to the particle number current, and hence we can write the effective action as a functional of $X$, valid for any trap. In general, the coupling to the trap is fixed by generalized coordinate invariance \cite{Son:2005rv}. The static superfluid background is $\phi_{\text{cl}}=\mu t$, where $\mu$ is the chemical potential,  and the expectation value of $X$ is $\langle X\rangle=\mu-V(x)$.

The effective action at low energies is a functional of $X$ with no additional derivatives,
 \begin{equation}
 \label{Eofstate}
    \begin{aligned}
S_\text{EFT}=\int dtd^2x P(X)~.
    \end{aligned}
\end{equation}
The effective theory for the fluctuations is obtained by expanding $P(X)$ around the classical background $\langle X\rangle$.   $P(X)$ is identified as the thermodynamic pressure at the chemical potential $\langle X\rangle$.  The particle number and current are given by $\rho=P'(X)$ and $\vec{J}=-P'(X)(\vec\nabla \phi)/m$.

In general, the form of $P(X)$ depends on the UV details of the superfluid, and it could be complicated even for weakly coupled microscopic models~\cite{Nicolis:2023pye}.  One well-studied UV completion is the Gross-Pitaevskii (GP) model in three-dimensional space, confined in a flat region of height $h$,
\begin{equation}\label{GP_action}
    \mathcal{L}_\text{GP}=\int_0^h dz\left[\Psi^*\left(i\partial_t+\frac{\nabla^2}{2m}-V\right)\Psi-\frac{\bar{g}}{4}|\Psi|^4\right]\,.
\end{equation}
The model~\eqref{GP_action} describes bosons with short-range repulsion in the s-wave. The field $\Psi$ marks the condensate wave function in the Hartree-Fock approximation, and the coupling is given by $\bar{g}=8\pi \ell_\text{s}/m>0$, where $\ell_\text{s}$ is the s-wave scattering length. The action~\eqref{Eofstate} is derived from~\eqref{GP_action} by setting $\Psi=e^{-i\phi}\sqrt{\rho/h}$, ignoring fluctuations in the $z$ direction, and working in the Thomas-Fermi approximation. This amounts to neglecting the Madelung quantum pressure term $\frac{1}{2m}\left(\nabla\sqrt{\rho}\right)^2$, which is justified when
\begin{equation}\label{eq_rho_condition}
    \left|\nabla\sqrt{\rho}\right|^2\ll g\rho^2\,,\qquad
    g\equiv\frac{m\bar{g}}{h}\,.
\end{equation}
Here we have introduced a dimensionless coupling $g$ for convenience. As long as the condition~\eqref{eq_rho_condition} holds, the equation of motion for the density field gives $\frac{g}{2 m}\rho\approx X$, and plugging back it into the action, we obtain the equation of state:
\begin{equation}\label{EoS_GP}
P(X)=\frac{m}{g}X^2\,.
\end{equation}
% where $g=m\,\bar{g}/h$. 
%\GC{This simple derivation shows that the effective theory holds when we coarse grain over distances much larger than the ``healing length'' $\xi^{-2}=g\rho$, which thus provides the cutoff~\eqref{EoS_GP}.}
This simple derivation shows that the effective theory holds when we coarse grain over distances much larger than the healing length $\xi^{-2}=mX$, which thus provides the cutoff to \eqref{Eofstate}.

In the following, we will often use the equation of state~\eqref{EoS_GP} as a benchmark for our results. Note that the equation of state~\eqref{EoS_GP}, and, more generally, the equation of state $P(X)\propto X^{1+d/2}$ in $d$-dimensional space, describes a system invariant under the non-relativistic conformal group \cite{Nishida:2007pj}; e.g., for $d=3$, this setup describes the finite density (zero temperature) phase of fermions at unitarity~\cite{2006PhRvA..74e3604W, Nishida:2010tm}. Conformal superfluids recently received much attention in the context of the large charge expansion, in both relativistic~\cite{Hellerman:2015nra, Monin:2016jmo, Cuomo:2022kio} and nonrelativistic~\cite{Kravec:2018qnu, Kravec:2019djc, Hellerman:2021qzz} contexts. 

The equation of motion that follows from eq.~\eqref{Eofstate} reads 
\begin{equation}
\label{EOM}
 m\partial_t P'(X) -\nabla [P'(X)(\nabla \phi)]=0~.
\end{equation}
The equation of motion always admits classical solutions $\phi_\text{cl}=\mu t$, where $\mu$ is the chemical potential.  Since $\rho=P'(x)$, the chemical potential $\mu$ controls the number of particles in the trap. Note that it is not physically meaningful to allow $P'(X)$ to attain negative values - this restricts the domain of integration in~\eqref{Eofstate} to the domain where $P'(X)$ is positive.  In general we expect $P'(X)= 0$ for $X=0$ \cite{Nicolis:2023pye}.

The equation of motion~\eqref{EOM} has to be supplemented by appropriate boundary conditions. To guarantee that particles do not flow through the boundary, to the leading order in derivatives we impose~\cite{Hellerman:2020eff, Cuomo:2021cnb}, 
\begin{equation}
\label{eq_boundary condition}
    \vec{J}\cdot  \hat n=0~,
\end{equation}
where $\hat{n}$ is transverse to the boundary of the superfluid. We remark that~\eqref{eq_boundary condition} is a universal \emph{effective} boundary condition of the low-energy theory. It was demonstrated in~\cite{Cuomo:2021cnb} that the Neumann condition~\eqref{eq_boundary condition} correctly describes the phase-shift of low-energy phonons scattering off a boundary, even when the microscopic model is supplemented with Dirichlet boundary conditions for the density field $\rho=0$. As we verify below, \eqref{eq_boundary condition} applies as long as the scales we discuss are within the EFT cutoff.

Finally, expanding around the static solution $\phi_\text{cl}=\mu t$ and neglecting the trapping potential, we find that low-energy phonons have the dispersion relation $\omega\sim c_{\text{s}}|\vec{k}|$, where the speed of sound is given by 
\begin{equation}\label{eq_cs2}
c_{\text{s}}^2=\frac{P'(\mu)}{m P''(\mu)}\,.
\end{equation}

\section{The Giant Vortex}

When the trap is axisymmetric $V(x)=V(r)$, there is another set of solutions to~\eqref{EOM}: 
\begin{equation}
\label{eq_GV classical profile}
    \phi_{\text{GV}}=\bar{\mu} t-L\theta \,,
\end{equation}
where $L\in\mathbb{Z}$ is the vorticity and $\mub$ should not be confused with the chemical potential. We assume for definiteness $L>0$. When $L\sim O(1)$, this solution describes a microscopic vortex and can be analyzed within the formalism of~\cite{Horn:2015zna}. Here we are interested in the giant vortex limit $L\gg 1$, in which the vortex core is macroscopic. The properties of such solutions and fluctuations about these solutions are the main subject of this paper. Recently, vortices with large winding numbers were also analyzed in relativistic superfluids \cite{Cuomo:2022kio, Kourkoulou:2023xqe} and superconductors \cite{Evans:2020uui,Penin:2020cxj,Gates:2022bnv}.

Let us consider first the giant vortex in a hard trap which is a cylinder of radius $R$, and assume the equation of state~\eqref{EoS_GP}. We denote with $\Omega=\frac{L}{m R^2}$ the angular velocity near the edge of the trap. Since
\begin{equation}
    X=\mub-{L^2\over 2mr^2}=\mub-\frac{m\Omega^2 R^4}{2r^2}\,,
\end{equation}
the density is non-negative in the domain $R\geq r\geq R_{\text{GV}}$, with $R_{\text{GV}}^2=\frac{m  \Omega ^2 R^4}{2 \mu }$. The superfluid occupies an annulus and spins with axial superfluid velocity $v(r)=\Omega\frac{R^2}{r}$.
We can relate $\mub$ to the number of particles by integrating the density,  leading to 
\begin{equation}\label{N_GP}
N=\frac{\pi  }{g}  (m R^2\Omega)^2  \left( {R^2\over R_{\text{GV}}^2}-2\ln{R\over R_{\text{GV}}}-1\right)
\end{equation}
where we solved for $\mub$ in terms of $R_{\text{GV}}$. Equation \eqref{N_GP} allows us to compute the radius of the giant vortex $R_{\text{GV}}$ in terms of the vorticity and the parameters of the trap and superfluid.  If $R_{\text{GV}}$ is not too close to $R$, we will obtain approximately 
$R_{\text{GV}}\sim L\xi$ in terms of the ``healing length'' $\xi^{-2}=g\rho$. We see that we need a large angular velocity $L\gg 1$ to create a macroscopic ($R_{\text{GV}}\gg \xi$) hole. When $R_{\text{GV}}\ll R$, we will obtain approximately 
$R_{\text{GV}}\sim L\xi$ in terms of the healing length. Therefore, a large angular velocity $L=m \Omega R^2\gg 1$ is required to create a macroscopic fluid ($R_{\text{GV}}\gg \xi$). For a narrow GV annulus where $R_{\text{GV}}=R-\delta$ with $\delta\ll  R$, we find:
\begin{equation}\label{eq_pre_N}
N\simeq \frac{2 \pi }{g} (m R\delta \Omega)^2\,. 
\end{equation}

Let us comment on the EFT validity in a narrow GV annulus. The centrifugal force guarantees the absence of light modes in the vortex core. In the superfluid annulus, we may reliably work purely in terms of the phase field as long as condition~\eqref{eq_rho_condition} holds. Around the intersection at $r=R_{\text{GV}}$, the equation of motion states that the Thomas-Fermi approximation breaks down at $r=R_{\text{GV}}+\epsilon$, where
\begin{equation}\label{eq_delta_r}
    \epsilon\simeq (m^2 R \Omega^2)^{-\frac{1}{3}}\,,
\end{equation}
and $\epsilon$ should be understood as the thickness of the boundary layer. It is important to distinguish $\epsilon$ from the width of the GV annulus $\delta$, which scales as
\begin{equation}\label{eq_delta}
\delta\simeq \frac{\bar\mu}{m R\Omega^2}-\frac{R}{2}\simeq (m^2 \xi^2\Omega^2   R)^{-1}
\end{equation}
where $\xi$ is the healing length measured near the peak of the particle density at $r=R$.

We can safely replace the boundary layer where the EFT breaks down with an effective boundary \eqref{eq_boundary condition} as long as its thickness is much smaller than the annulus size. From Eq.s~\eqref{eq_delta_r} and~\eqref{eq_delta} we conclude that this is justified when
\begin{equation}\label{eq_pre_condition}
\frac{\epsilon}{\delta}\simeq \left(\frac{\xi}{\delta}\right)^{\frac{2}{3}}\ll 1\,.
\end{equation}
Equation \eqref{eq_pre_condition} states that for the EFT to be valid, the GV annulus must be wide when compared to the healing length measured around the peak of the particle density. Equivalently, the condition requires that $ m R^2\Omega\ll (gN)^{3/2}$.

A similar discussion holds for systems with a different equation of state and in generic traps. In all cases, we find that the EFT holds as long as the GV scales are larger than the healing length measured near the point where the density peaked. Using the equation of state, this condition can be expressed in terms of measurable parameters on a case-by-case basis. For instance, for the quadratic model~\eqref{EoS_GP} in a simple power law trap $V(r)=\frac{\omega}{2q}\left( m\omega r^2\right)^q$ (with $q>1$), we find that the effective theory holds as long as $\left(\Omega/\omega\right)^{\frac{2(q+1)}{3(q-1)}}\ll (g N)^{4/3}$.

In this work, the stability of the solution~\eqref{eq_GV classical profile} at a fixed particle number and angular momentum is not discussed. This question was partially studied in previous works~\cite{PhysRev.153.285,fetter2005rapid,correggi2013vortex}, and it is generally believed that the superfluid eventually settles in a giant vortex state as the trap rotation speed is increased.\footnote{Here we study only single species superfluids. Recently, \cite{2023PhRvA.107e3317R} proposed a mechanism to stabilize vortices with $L>1$ in multi-species condensates.} Our discussion will apply to systems at a sufficiently large rotation frequency, and we leave the detailed study of the phase transition to the GV state to a future work~\cite{future}.

\section{Fluctuation Spectrum: Gross-Pitaevskii Model in a Hard Trap}

We now study the fluctuations of the giant vortex in a hard trap with the equation of state~\eqref{EoS_GP}. 
An axially symmetric trap explicitly breaks boosts and translations.\footnote{There is a well-known exception, the quadratic trap, which admits an extended symmetry group equivalent to that of a particle in a magnetic field~\cite{Gibbons:2010fb}; this symmetry group is important for the existence of an emergent translational symmetry group in the vortex lattice~\cite{Moroz:2018noc}.} Additionally, the solution~\eqref{eq_GV classical profile} spontaneously breaks the $U(1)$ particle number $N$, time translations $H$ and rotations $J$ down to the two linear combinations $H-\mub N$ and $J-L N$. The existence of these two unbroken generators allows us to organize fluctuations into modes with well-defined frequencies and angular momentum. In contrast, the vortex lattice does not admit an unbroken rotation generator.

We denote the fluctuation field $\varphi \in \mathbb{S}^1$ and write $\phi=\phi_{\text{GV}}+\varphi$. We will assume that $\varphi$ does not wind around the $\theta$-coordinate to avoid double counting of the modes, which is justified for a thermodynamically stable state. The fluctuation Lagrangian to quadratic order reads
\begin{equation}
\label{eq_hard disk complete action}
    \begin{aligned}
\mathcal{L}_{\text{flu}}=&\frac{m}{g} \left(\partial_t\varphi+\frac{\Omega R^2 }{r^2}\partial_\theta \varphi \right)^2\\
 &-\frac{m  \Omega^2 R^4}{2 g}\left({1\over R_{\text{GV}}^2}-{1\over r^2}\right)(\nabla\varphi)^2\,.
    \end{aligned}
\end{equation}
Equation \eqref{eq_hard disk complete action} is supplemented with \emph{effective} boundary conditions at the hard wall $r=R$ and at $r=R_{\text{GV}}$, where the Thomas-Fermi approximation breaks down. Note that we have neglected the boundary layer thickness $\epsilon$, and as we have commented, eq.~\eqref{eq_pre_condition} is the unique boundary condition compatible with low-energy symmetries of the system ~\cite{Cuomo:2021cnb}. We conclude that the boundary condition reads
\begin{equation}
    \left({1\over R_{\text{GV}}^2}-{1\over r^2}\right)\pd_r \varphi\vert_{r=R,R_{GV}}=0\,,
\end{equation}
which is the Neumann condition at $r=R$ and demands regularity of $\varphi$ at $r=R_{GV}$.

We study the fluctuations with the ansatz $\varphi=e^{-i\omega t+in\theta}Y(r)$. The problem simplifies if we denote $R_{\text{GV}}/R=\sqrt{\lambda/(\lambda+1)}$, with $\lambda\in \mathbb{R}^+$, such that $\lambda\ll 1$ is the limit where the GV hole is small compared to the disk size while $\lambda \gg 1$ is the limit of a narrow GV. Furthermore, we introduce a new coordinate $z\in \mathbb{R}^+$ and let $r/R=\sqrt{(\lambda+e^{-z})/(\lambda+1)}$. The equation of motion in terms of $Y$ reduces to a Schr\"odinger problem,
\begin{equation}
-\partial_z^2Y+\mathcal{V}(z) (Y/4)=0\,,
\end{equation}
where
\begin{equation}
\label{eq_Schrodinger potential}
    \begin{aligned}
\mathcal{V}(z)=\left(\frac{n}{ \lambda e^z+1}\right)^2-\frac{2\lambda}{e^z} \left(\frac{\omega }{\Omega(\lambda +1)}-\frac{n e^z}{\lambda  e^z+1}\right)^2\,,
    \end{aligned}
\end{equation}
and the boundary condition reads $\partial_z Y=0$ at $z=0$ as well as $z=+\infty$. 

At the inner edge of the GV, $z\gg 1$, the  potential is
\begin{equation}\label{ExpNear}
    \begin{aligned}
\mathcal{V}(z)=-\frac{2}{\lambda}\left[\frac{ \omega  R_\text{GV}^2}{\Omega R^2}-n\right]^2e^{-z}+O\left(e^{-2z}\right)\,.
    \end{aligned}
\end{equation}
The potential is always attractive (negative), unless $\frac{ \omega R_\text{GV}^2}{\Omega R^2}-n=0$.  This is a striking feature: Despite the fast swirling of the superfluid,  one finds that some phonons are attracted to the inner edge of the GV.

We discuss in detail only the narrow limit $\lambda\rightarrow\infty$. 
Then we find the radial wave functions $Y=J_0\big(k e^{-z/2}\big)$ with wave number 
$k=\sqrt{2/\lambda}|\omega R_\text{GV}^2/(\Omega R^2)-n|$. 
$k$ is quantized by virtue of the boundary conditions at the hard trap. We find that either $k=0$ or $k_{n'}=j_{1,{n}'}$, where $j_{a,b}$ is the $b$-th zero of the Bessel function $J_a$.
The modes with a nontrivial radial profile $n'>0$ lead to the dispersion relation 
\begin{equation}
\label{eq_hard disk dispersion general}
    \begin{aligned}
\omega_{n,{n}'}=\Omega\left[n+\sqrt{\frac{R}{\delta}}\frac{j_{1,{n}'}}{2}+O\left(\sqrt{\frac{\delta}{R}}\right)\right]\,,
    \end{aligned}
\end{equation}
Due to the nontrivial profile over the annulus width, eq.~\eqref{eq_hard disk dispersion general} yields a large gap $\omega-\Omega n\sim \Omega\sqrt{R/\delta}$ in the rotating frame. 

The states $k=0$ are much more interesting. We find a mode with profile $Y=1+O(\delta^2/R^2)$ and dispersion 
\begin{equation}
\label{eq_hard disk dispersion chiral}
    \begin{aligned}
\omega_{n,0}=\Omega\left[n+\sqrt{\frac{\delta }{2R}}|n|+O\left(\frac{\delta}{R}\right)\right]~.
    \end{aligned}
\end{equation}
Here we included the small correction $\sqrt{\frac{\delta }{2R}}|n|$ compared to~\eqref{intrdis}. In the limit $\delta/R\rightarrow 0$, the wave profiles of these states are co-moving with the trap. We refer to such modes as chiral modes; note that their gap in the rotating frame $\omega-\Omega n$ is much smaller than the angular velocity. See Fig.~\ref{CandG} for a graphical summary of the spectrum in the hard trap. Our derivation remains valid as long as the condition~\eqref{eq_rho_condition} is satisfied everywhere but in the small region $\epsilon\ll \delta$. This implies that the dispersion relations~\eqref{eq_hard disk dispersion general} and~\eqref{eq_hard disk dispersion chiral} hold only for 
sufficiently long wavelengths, where $|n|/R\ll \xi^{-1}$ and $|n'|/\delta\ll \xi^{-1}$.

A few remarks are in order. First, note that the chemical potential is $\mu\simeq \mub-\Omega L$ in the narrow limit. The system admits the unbroken Hamiltonian $H+\Omega J-\mu N$, and chiral modes are the lowest-lying excitations of it. 

Second, at a fixed particle number $N$, the angular velocity $\Omega$ and the annulus width $\delta$ are related through the equation of state $P(X)$. In the GP model, we can use relation~\eqref{eq_pre_N} to express the results~\eqref{eq_hard disk dispersion general} and~\eqref{eq_hard disk dispersion chiral} in terms of other physical quantities. 

Finally, even though the particle density peaks near the wall at $r=R$, the modes we discussed are approximately delocalized over the full GV annulus. This should be contrasted with several instances of chiral edge modes in classical and quantum fluids. A well-known example is that of coastal Kelvin waves in fluid mechanics, where chiral edge modes near the coast arise due to the Coriolis force \cite{vallis_2006, Tong:2022gpg,2023arXiv230305669M}. Another example is the chiral motion of waves at the edge of a superfluid vortex bundle in a cylindrical container \cite{sonin2016dynamics}. In both examples, the corresponding modes are localized near the edge of the region occupied by the fluid.

\section{Fluctuation Spectrum: Generic \texorpdfstring{$P(X)$}{P(X)} Model in a Hard Trap}

We now generalize the discussion in the last section to an arbitrary equation of state $P(X)$. We focus directly on the experimentally relevant narrow limit. We linearize $\langle X\rangle$ around the edge $r=R$, at which it is peaked:
\begin{equation}\label{eq_X_trap_linear}
\langle X\rangle=\mu_{\text{eff}} (1+y)\left [1-\frac{3\delta}{R}y+O\left(\frac{\delta^2}{R^2}\right)\right] \,,
\end{equation}
where $\delta=R-R_{\text{GV}}$ is the thickness of the annulus, $\mu_{\text{eff}}=\mub-m\Omega^2R^2/2$, and $y=(r-R)/\delta\in [-1,0]$.  We imposed that $\langle X\rangle$ vanishes at the inner edge $y= -1$. 

We analyze the spectrum of fluctuations in a series expansion for $\delta/R\ll 1$. This amounts to an expansion of the equations of motion analogously to what we did in~\eqref{eq_X_trap_linear}. The details are given in the Appendix.
For the modes with a nontrivial profile in the radial direction we find 
\begin{equation}\label{eq_hard_trap_ds1}
\omega_{n,n'}=\Omega\left[n+\sqrt{\frac{R}{\delta}}k_{n'}+O\left(\sqrt{\frac{\delta}{R}}\right)\right]\,,
\end{equation}
where $k_{n'}$ is an $O(1)$ number which depends upon the equation of state. For the quadratic model~\eqref{EoS_GP} we found in eq.~\eqref{eq_hard disk dispersion general} $k_{n'}=j_{1,n'}/2$.

The $n'=0$ chiral modes in eq.~\eqref{eq_hard_trap_ds1} again require separate treatment and a computation of subleading corrections. 
We arrive at the final result
\begin{equation}\label{eq_hard_chiral_NLO}
\omega_{n,0}=\Omega\left[n+\alpha_P\sqrt{\frac{\delta}{R}}|n|+O\left(\frac{\delta}{R}\right)\right]
\end{equation}
where $\alpha_P$ is an $O(1)$ coefficient given by
\begin{equation}\label{eq_ap_hard}
\begin{aligned}
\alpha_P & =\sqrt{\frac{\int  d^2x {P}'\left(\langle X\rangle\right)}{\mu_{\text{eff}}\int  d^2x {P}''\left(\langle X\rangle\right)}}>0\,.
\end{aligned}
\end{equation}
In eq.~\eqref{eq_ap_hard} $\langle X \rangle$ is evaluated to its leading order in $\delta/R$. 
% Physically, $\alpha_P$ is an averaged sound speed in units of $\sqrt{\mu_\text{eff}/m}$. 
The term proportional to $\alpha_P$ in eq.~\eqref{eq_hard_chiral_NLO} lifts the pathological degeneracy of the ground state. 

The chiral mode dispersion relation~\eqref{eq_hard_chiral_NLO} has a clear physical interpretation. Upon a comparison of Eqs.~\eqref{eq_cs2} and~\eqref{eq_ap_hard}, $\alpha_P \equiv \sqrt{m/\mu_{eff}}c_{\text{s,eff}}=O(1)$ can be understood as an averaged sound speed in proper units. We notice that eq.~\eqref{eq_X_trap_linear} implies
 \begin{equation}\label{eq_hard_scaling}
 \Omega^2\simeq\frac{\mu_{\text{eff}}}{m\delta R}\sim \frac{c_{\text{s,eff}}^2}{\delta R}\,.
\end{equation}
Since the momentum of a mode delocalized over the annulus is $\sim 1/R$, we expect the energy of chiral modes in the rotating frame to scale as $\omega-\Omega n\sim c_{\text{s,eff}}/R\sim \Omega\sqrt{\delta/R}$. This estimate agrees with the explicit result~\eqref{eq_hard_chiral_NLO}. Similar considerations justify the scaling $\omega-\Omega n\sim c_{\text{s,eff}}/\delta\sim \Omega\sqrt{R/\delta}$ for the non-chiral modes in eq.~\eqref{eq_hard_trap_ds1}. 

However, the exact form of the dispersion relation and the value of $\alpha_P$, cannot be obtained from dimensional considerations. Indeed, such quantities depend on the nontrivial profile of the superfluid; see the Appendix for details and examples.

\section{Fluctuation Spectrum: Smooth Traps}

We finally discuss the narrow GV in a smooth trap $V(r)$. The superfluid resides between the zeros of the density $P'(X)$. We assume that they coincide with the zeros of $X=\mub-V(r)-{L^2\over 2mr^2}$. We define $R$ to be the point at which $\langle X\rangle$ reaches its maximum; i.e., we have
\begin{equation}\label{eq_smooth_R_def}
0=\frac{L^2}{m R^3}-V'(R)=m \Omega^2 R-V'(R)\,,
\end{equation}
where in the last equality $\Omega=L/(m R^2)$ denotes the angular velocity at the point $r=R$. 
We also assume that~\eqref{eq_smooth_R_def} admits a unique solution. This is, for instance, the case for the power law trap.

In the narrow limit, it is convenient to expand $\langle X\rangle$ near its maximum, similar to eq.~\eqref{eq_X_trap_linear}. Because of eq.~\eqref{eq_smooth_R_def} the expansion starts at quadratic order:
\begin{equation}\label{eq_smooth_X_exp}
\langle X\rangle=\mu_{\text{eff}}\left[1-y^2+c_3\frac{\delta}{R}y^3+O\left(\frac{\delta^2}{R^2}\right)\right]\,,
\end{equation}
where $\mu_{\text{eff}} =\mub-\frac12 M\Omega^2 R^2-V(R)$,  $\mu_{\text{eff}}/\delta^2=\frac12V''(R)+\frac{3}{2}M\Omega^2$, and we defined a new variable, $y=(r-R)/\delta$.  Therefore, $R$ approximately coincides with the center of the annulus, and the superfluid density vanishes at $y \simeq\pm 1$. This approximation is accurate as long as the potential $V(r)$ is not too steep; e.g., for $V(r)\sim r^q$ we find $q\ll R/\delta$.

The calculation of the spectrum proceeds analogously to that in the previous section. It turns out that in the narrow limit, the information about the trapping potential is contained in the following factor
\begin{equation}
\gamma_V\equiv \sqrt{\frac{\mu_\text{eff}}{ 2m}}\frac{1}{\delta \Omega}=O(1)\, . 
\end{equation}
$\gamma_V$ roughly characterizes the steepness of the trap around $r=R$. For example in a power-law trap, $V\sim r^q$ reads $\gamma_V=\sqrt{(q+2)/4}$. We obtain the following spectrum:
\begin{equation}
\label{eq_smooth general dispersion}
\omega_{n,n'}=\Omega \left[n+\sqrt{2} \gamma_V k_{n'}+O\left(\frac{\delta}{R}\right)\right]\,.
\end{equation}
The $n'=0$ solutions have $k_{0}=0$ and $Y=\text{const}.$: These are the chiral modes which we will discuss separately. 
 Remarkably, we also find another set of solutions, $n'=1$, for which we universally have $k_{1}=\sqrt{2}$. (These can be interpreted as approximate gapped Goldstone modes; see the Appendix for details.)
 The solutions $n'>1$ depend upon the equation of state and have $ k_{n}'>\sqrt{2}$, with frequencies $\sim c_{\text{s,eff}} /\delta$. As an example, in the model~\eqref{EoS_GP} we find
$k_{n'}=\sqrt{n'(n'+1)}$. In the Appendix, explicit results for an arbitrary power-law equation of state $P(X)\propto X^q$ are also computed.

The calculation of the subleading correction to the frequency of the chiral modes again proceeds analogously.  
Interestingly, they are non-tachyonic only as long as $\gamma_V>1$, i.e. as long as the trap is steep enough. For instance, a power-law trap needs to be steeper than quadratic. Notice that the same is true for the vortex lattice. Physically, this is because we need the potential to balance the centrifugal force.

\begin{figure}[!h]
\includegraphics[width=0.51\textwidth ]{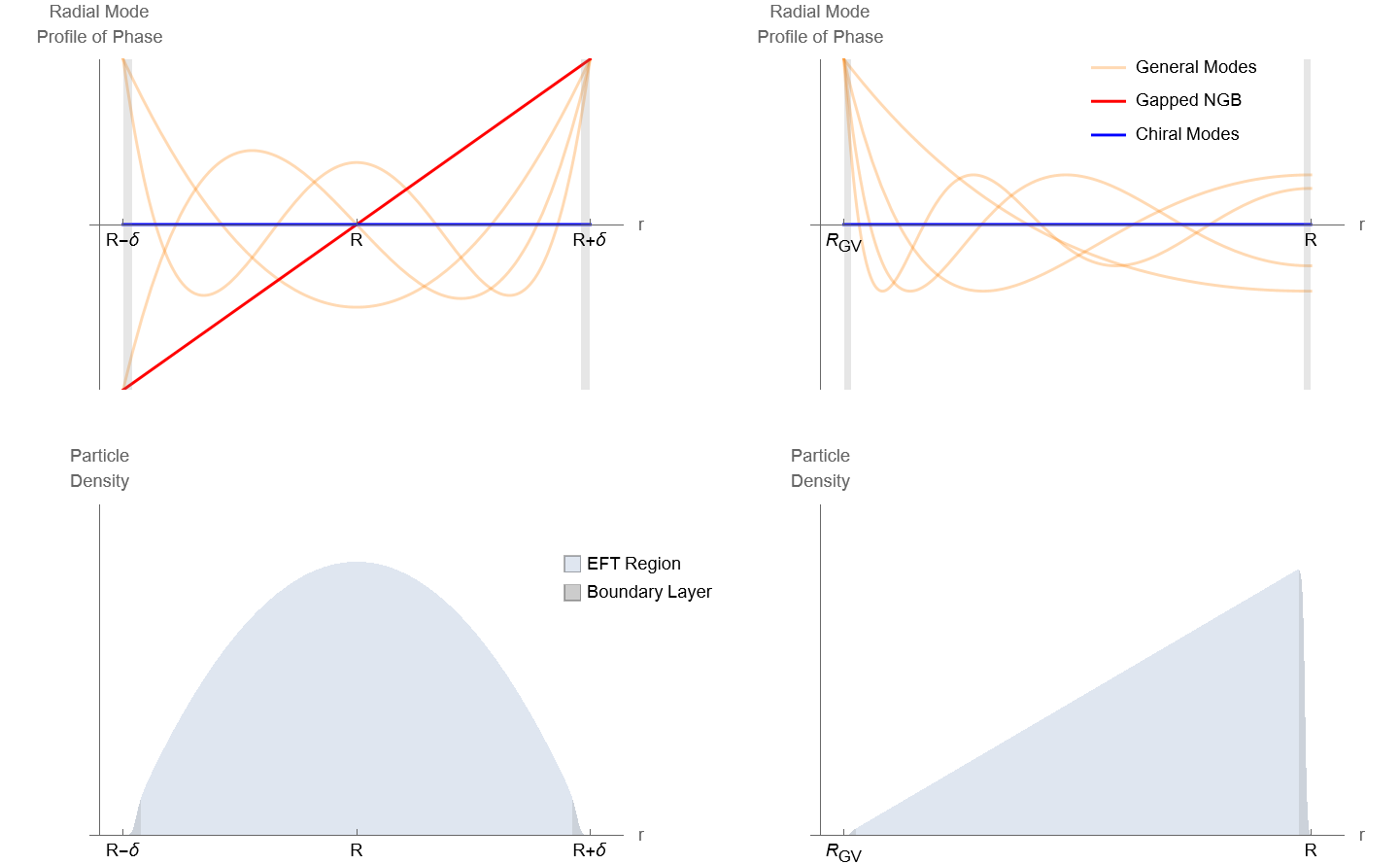}
\caption{ A radial section of the excitation mode profile (top) and the superfluid density profile and density excitations (bottom). The GP model is schematically plotted for a small trap (left) and a hard trap (right). }
\label{CandG}
\end{figure}

The result for the frequency of the chiral modes at subleading order reads
\begin{equation}\label{eq_chiral_NLO_smooth}
\omega_{n,0}=\Omega \left[n+\alpha_P\sqrt{2(\gamma_V^2-1)}|n|\frac{\delta}{R}+O\left(\frac{\delta^2}{R^2}\right)\right]\,,
\end{equation}
where $\alpha_P$ is an $O(1)$ number defined as in \eqref{eq_ap_hard}. See Fig.~\ref{CandG} for a qualitative summary of the spectrum of the GV in both the hard and smooth traps.

To understand the result~\eqref{eq_chiral_NLO_smooth}, we note that the fact that the expansion in eq.~\eqref{eq_smooth_X_exp} starts at quadratic order implies an important difference with respect to the hard trap. For a sufficiently smooth trap,  we expect the scaling $\partial^n_rV(r)\sim m\Omega^2 R^{2-n}$. From the definition of $\delta$ below eq.~\eqref{eq_smooth_X_exp} we infer that
\begin{equation}\label{eq_smooth_scaling}
\Omega^2\sim\frac{\mu_{\text{eff}}}{m\delta^2}\sim\frac{c^2_{\text{s,eff}}}{\delta^2}\,,
\end{equation}
where we again used the estimate $c_{s,\text{eff}}^2\equiv\frac{\mu_{eff}}{m}\alpha_P^2\sim \frac{\mu_{\text{eff}}}{m}$. The scaling~\eqref{eq_smooth_scaling} differs by a factor $R/\delta$ compared to eq.~\eqref{eq_hard_scaling}. Therefore, we expect chiral modes in a smooth trap to have gaps in the rotating frame that scale as $c_{\text{s,eff}}/R\sim \Omega\delta/R$, which is small compared to ~\eqref{eq_hard_chiral_NLO} for the hard trap. That is, indeed, what we find in eq.~\eqref{eq_chiral_NLO_smooth} for $\alpha_P\sim \gamma_V\sim O(1)$.

Finally, we remark that physical quantities are related through the equation of state $P(X)$. As an example of the GP model~\eqref{EoS_GP} in a power law trap $V(r)=\frac{ \varpi }{q}(m \varpi  r^2)^{\frac{q}{2}}$ with $q>1$, we find:
\begin{align}
R&=\frac{1}{\sqrt{m  \varpi }}\left(\frac{\Omega }{ \varpi }\right)^{\frac{2}{q-2}}\,,\\
\delta&=
\frac{1}{\sqrt{m  \varpi  }}
\left[\frac{3 g^2 N}{8\pi (q+2)}\right]^{\frac{1}{3}} \left(\frac{\Omega }{ \varpi }\right)^{-\frac{2 (q-1)}{3 (q-2)}}\,.
\end{align}
Note that for $q=4$ the area of the annulus $A\simeq 2\pi\delta R$ is independent of the rotation speed $\Omega$, as noted in \cite{fetter2005rapid}, and more generally we find $A\propto (\Omega/\varpi)^{-\frac{2(q-4)}{3(q-2)}}$.
These equations, together with the relations
\begin{equation}
    \alpha_P=\sqrt{\frac{2}{3}}\,,\quad
    \gamma_V=\sqrt{\frac{q+2}{4}}\,,
\end{equation}
allow us to express the results~\eqref{eq_smooth general dispersion} and~\eqref{eq_chiral_NLO_smooth} in terms of $q$, $g$, $\Omega$ and the particle number $N$.

\section{EFT of the Narrow Giant Vortex}\label{sec_EFT_narrow}

Here we discuss the EFT of the superfluid phase that leads to the peculiar infinite degeneracy of the ground state at fixed angular momentum.
The leading order theory from which~\eqref{intrdis} follows is 
\begin{equation}
\label{WCFT}
S\sim \int dtd\theta \left(\partial_t\varphi+\Omega\partial_\theta \varphi \right)^2 \sim \int dx^+dx^-(\partial_+\varphi)^2~.
\end{equation}
where we switched to the coordinates $x^{\pm}=t\pm mR^2\theta/L$. This action has chiral conformal symmetry 
$x^-\to f(x^-)$ with $\varphi\to \varphi/\sqrt{f'(x^-)}$ as well as chiral translations $x^+\to x^++g(x^-)$. These transformations define the 
warped conformal group (see, e.g.~\cite{Jensen:2017tnb}).  Additionally, the action~\eqref{WCFT} admits the \emph{fractonic} shift symmetry $\varphi\rightarrow \varphi+f(x^-)$. This symmetry group implies the existence of infinitely many chiral solutions with zero energy and it is therefore responsible for the enormous ground state degeneracy. Corrections to~\eqref{WCFT} lift this degeneracy.  From the perspective of the effective theory~\eqref{WCFT}, corrections arise from the term $\sim \frac{c_{\text{s,eff}}^2}{R^2} \int dx^+dx^- (\partial_\theta\varphi)^2$ and lead to the required modification in the dispersion relation.

Additional small corrections which we compute in the Appendix for smooth traps remove the remaining degeneracy in excited states. The degeneracy among excited states is split due to a higher derivative term such as $\sim  \int dx^+dx^- (\partial^2_\theta\varphi)^2$ as we illustrate in the Appendix.

\section{Discussion and Outlook}

In this paper, we discussed the fluctuations of superfluid giant vortices. Our main results are the dispersion relations of the chiral modes, eq.~\eqref{eq_hard_chiral_NLO} in the hard trap and eq.~\eqref{eq_chiral_NLO_smooth} in a smooth trap. These results bear the most generality of a large collection of physical systems, with $O(1)$ parameters $\alpha_P$ and $\gamma_V$, which we explicitly calculated.

We did not investigate the thermodynamical stability of the GV ground state, and thus the lower critical value of $\Omega$ for our results to hold remains unspecified. Former analyses~\cite{PhysRev.153.285} suggest that the superfluid flow becomes irrotational for $m\Omega\delta^2\ln(\delta/\xi)\lesssim 1$. For the GP model in a hard trap, we conclude that our result should apply for $\Omega\gtrsim\frac{g N}{2\pi m R^2}$ from eq.~\eqref{eq_pre_N}.

In practice we expect our results to hold also for lower values of $\Omega$, such that the flow is not completely irrotational and a few vortices are present. A recent experiment~\cite{guo2020supersonic} succeeded in creating a stable rotating superfluid ring of ${}^{87}\text{Rb}$ condensate in an anharmonic trap. The achieved rotation speed suggests that the observed state did not yet reach the GV phase. Nonetheless, intriguingly, the same experiment observed shape deformations co-moving with the fluid. In the setup of \cite{guo2020supersonic}, shape deformations are created via time-dependent weak elliptical deformations of the trap. The experimental results suggest that the regime of applicability of our results might be larger than the naive window mentioned above. In the future, it would be interesting to perform similar experiments at larger rotation speeds and measure the dispersion relations that we studied in this work.

As already mentioned, it is interesting to revisit the transition from the vortex lattice state to a purely irrotational flow. Essential questions regarding this process include what the order of the phase transition is: first, second, or higher? How does the dispersion relation of the lowest lying mode (in the rotating frame) deform from the Tkachenko behavior to the linear ones predicted in this work? To the best of our knowledge, former works on the subject \cite{fischer2003vortex,2004PhRvA..69c3608A,fetter2005rapid,2005PhRvL..94j0402C,correggi2013vortex} have not yet settled these questions.

Finally, we remark that the one-dimensional effective theory~\eqref{WCFT} also describes the fluctuations of a GV in the relativistic context~\cite{Cuomo:2022kio}. Warped conformal symmetries have also appeared in the description of the near horizon of spinning black holes~\cite{Guica:2008mu, Hofman:2011zj}. It would be intriguing to explore the connection with the literature further.

\begin{acknowledgments}
\section{Acknowledgments}
We thank K. Jensen for useful discussions and A. Nicolis for useful discussions and collaboration on a related project. GC is supported by Simons Foundation Grant No. 994296 (Simons Collaboration on Confinement and QCD Strings). ZK and SZ are supported in part by Simons Foundation Grant No. 488657 (Simons Collaboration on the Non-Perturbative Bootstrap), BSF grant No. 2018204, and NSF Award No. 2310283.
\end{acknowledgments}

\onecolumngrid

\section*{Appendix}

With a general equation of state, the Lagrangian density for the fluctuations is
\begin{equation}\label{eq_fluct_act}
\mathcal{L}_{\text{flu}}=\frac{{P}''(\langle X\rangle)}{2}\left(\partial_t\varphi+\frac{L}{m r^2}\pd_{\theta}\varphi\right)^2-\frac{{P}'(\langle X\rangle)}{2m}(\nabla\varphi)^2\,.
\end{equation}
Formally, to leading order in $\delta/R$, the equations of motion take the same form for both the smooth and the hard traps. To see this we express~$\langle X\rangle=\langle X\rangle_0+O(\delta/R)$, where $\langle X\rangle_0$ is the leading term in Eq.s~\eqref{eq_X_trap_linear} and~\eqref{eq_smooth_X_exp}. We expand the terms in eq.~\eqref{eq_fluct_act} as
\begin{align}
&\frac{L}{m r^2}\pd_{\theta}\varphi=\Omega\left[\pd_\theta\varphi-2y\partial_\theta\varphi \frac{\delta}{R}+O\left(\frac{\delta^2}{R^2}\right)\right]\,,\\
&(\nabla\varphi)^2=\frac{1}{R^2} \left[\frac{R^2}{\delta^2}(\pd_y\varphi)^2+(\pd_\theta\varphi)^2+O\left(\frac{\delta}{R}\right)\right] \,
\end{align}
where $y$, $\delta$, and $R$ are defined below~\eqref{eq_X_trap_linear} for the hard trap, and below~\eqref{eq_smooth_X_exp} for a smooth trap.

It is then straightforward to derive the equations of motion and solve them perturbatively in $\delta/R$. We adopt the ansatz~$\varphi=e^{-i\omega t+in\theta}Y(r)$, and we find the following equation at leading order:
\begin{equation}\label{eq_radial equation}
m\delta^2(\omega-\Omega n)^2{P}''(\langle X\rangle_0)Y(y)+
\pd_y\left[{P}'(\langle X\rangle_0)Y'(y)\right]=0\,.
\end{equation}
This is a second order ordinary differential equation for $Y(y)$, subject to the boundary conditions \eqref{eq_boundary condition}, which explicitly read
\begin{equation}
\label{eq_deltaY_bc}
{P}'(\langle X\rangle_0)Y'(y)=0\,.
\end{equation}
at the boundary of the annulus.

\subsection{The hard trap}

In a hard trap, we have $\langle X\rangle_0=\mu_\text{eff}(1+y)$ and $\mu_\text{eff}=m\Omega^2 R \delta $. The superfluid annulus terminates at the inner edge $y=-1$ and the hard cutoff $y=0$. From the solution of eq. \eqref{eq_radial equation} we obtain the dispersion relation \eqref{eq_hard_trap_ds1}. We solved analytically for the wave functions $Y(y)$ and the values of $k_n$ for equations of state of the form $P(X)\propto X^q$ with $q>1$: 
\begin{equation}
\begin{aligned}
k_{{n}'}=&\frac{j_{q-1,{n}'}}{2\sqrt{q-1}}\,,\\
Y_{{n}'}=&(1+y)^{1-\frac{q}{2}}J_{q-2}\left(j_{q-1,{n}'}\sqrt{1+y}\right)\,.
\end{aligned}
\end{equation}
These modes, as shown in \eqref{eq_hard_trap_ds1}, are very heavy excitations.

To leading order in $\delta/R$, we also have solutions $\omega_{n,0}=\Omega n$ with $Y=1$. They correspond to the special chiral modes, and they require a separate treatment. We make the ansatz 
\begin{equation}\label{eq_app_ansatz}
\begin{aligned}
\omega_{n,0}=& \Omega\left[n+\sqrt{\frac{\delta}{R}}\tilde{\omega}_n+O\left(\frac{\delta}{R}\right)\right]\,,\\
Y=&1+\frac{\delta^2}{R^2}\Tilde{Y}(y)+O\left(\frac{\delta^{5/2}}{R^{5/2}}\right)\,,
\end{aligned}
\end{equation}
and we obtain the following equation for the subleading order
\begin{equation}\label{eq_hard_trap_sub_pre}
n^2{P}'\left(\langle X\rangle_0\right)-\tilde{\omega}^2_n\mu_{\text{eff}}{P}''\left(\langle X\rangle_0\right)
=\pd_y\left[{P}'\left(\langle X\rangle_0\right) \tilde{Y}'(y)\right]\,.
\end{equation}
To determine $\tilde{\omega}_n$ we then simply need to integrate eq.~\eqref{eq_hard_trap_sub_pre} between $y=0$ and $y=1$. The term on the right-hand side vanishes by the boundary condition~\eqref{eq_deltaY_bc} and we find~\eqref{eq_hard_chiral_NLO}.

\subsection{The smooth trap}

In a smooth trap, we have $\langle X\rangle_0=\mu_\text{eff}(1-y^2)$, and $\mu_\text{eff}=2m(\gamma_V\Omega\delta)^2$. In terms of $y$, the annulus (approximately) occupies the interval $y\in[-1,1]$. The general dispersion relation is as stated in \eqref{eq_smooth general dispersion}. For models where $P(X)\propto X^q$ with $q>1$ we find
\begin{equation}
\begin{aligned}
k_{{n}'}=&\sqrt{\frac{{n}'({n}'+2q-3)}{q-1}}\,,\\
Y_{{n}'}=&\left(\frac{1+y}{2}\right)^{2-q} \, _2F_1\left(2-{n}'-q,{n}'+q-1;q-1;\frac{1-y}{2}\right)\,.
\end{aligned}
\end{equation}
Note in particular that for any $q>1$ the ${n}'=1$ mode universally yields $k_1=\sqrt{2}$ and $Y_1=y$. In other words, the solution ${n}'=1$ is independent of the equation of state, to leading order in $\delta/R$. This is because $Y= y$ corresponds to the action of a symmetry generator on the background~\eqref{eq_GV classical profile} for one of the extended symmetries of the quadratic trap $V(r)\sim r^2$~\cite{Gibbons:2010fb}, and to leading order in $\delta/R$ we cannot distinguish between different traps $V(r)$ in the expansion~\eqref{eq_smooth_X_exp}. This argument also fixes the gap of this mode,\footnote{For the quadratic trap the rotation speed coincides with the frequency of trap $V=\frac12m\Omega^2 r^2$, as well as $2\Omega=\sqrt{\mu_{\text{eff}}/m}/\delta$. The algebra implies the existence of a mode with angular momentum $n=1$ and gap $\omega-\Omega =2\Omega$, which coincides with our result.}  which is a gapped Goldstone since the associated generator does not commute with the unbroken Hamiltonian~\cite{Nicolis:2012vf,Watanabe:2013uya}. 

The calculation of the subleading correction to the frequency of the chiral modes proceeds analogously to that for the hard trap. We use the ansatz 
\begin{equation}
\begin{aligned}
\omega_{n,0}=& \Omega\left[n+\frac{\delta}{R}\tilde{\omega}_n+O\left(\frac{\delta^2}{R^2}\right)\right]\,,\\
Y=&1+\frac{\delta^2}{R^2}\Tilde{Y}(y)+O\left(\frac{\delta^{3}}{R^{3}}\right)\,,
\end{aligned}
\end{equation}
and we obtain the following equation at subleading order:
\begin{equation}\label{eq_smooth_trap_sub_pre}
n^2{P}'\left(\langle X\rangle_0\right)-\frac{\mu_\text{eff}}{2\gamma_V^2}(\tilde{\omega}_n+\sqrt{2}n y)^2{P}''\left(\langle X\rangle_0\right)
=\pd_y\left[{P}'\left(\langle X\rangle_0\right) \tilde{Y}'(y)\right]\,.
\end{equation}
The result is given in eq.~\eqref{eq_chiral_NLO_smooth} in the main text.

Finally,  eq.~\eqref{eq_chiral_NLO_smooth} implies that certain multi-phonon states are degenerate. This is the case for any two Fock space states of the same angular momentum $J=\sum_j n_j=\sum_l n_l$ where $n_j, n_l$ are either all positive or all negative. Given two such states, we find that the degeneracy between them is lifted at order $O(c_{\text{s,eff}}\delta^2/R^3)$.\footnote{To obtain this correction we computed the $O(\delta^4/R^4)$ term in the expansion of the wavefunction $Y(y)$ for the chiral modes.} The result reads:
\begin{equation}\label{eq_lift_gap_NNNLO}
E_{\{n_j\}}-E_{\{n_l\}}=\frac{\Omega\delta^3}{R^3}\left[
\beta_{P,V}\left(\sum_J|n_j|^3-\sum_l|n_l|^3\right)+O\left(\frac{\delta}{R}\right)\right]~,
\end{equation}
The coefficient $\beta_{P,V}=O(1)$ depends on both the equation of state $P(X)$ and the geometry of the trap $V$. It is given by
\begin{equation}
\label{C_3-def}
\beta_{P,V}=-\frac{\displaystyle \int dy\frac{\mu_{\text{eff}}}{P'(\langle X\rangle_0)}\left\{\int_{-1}^y \frac{d {y}'}{\gamma_V}\left[\frac{\gamma_V^2P'(\langle X\rangle_0)}{\mu_\text{eff}}-\left(\alpha_P\sqrt{\gamma_V^2-1}+\sqrt{2} {y}'\right)^2P''(\langle X\rangle_0)\right]\right\}^2}{\displaystyle\alpha_P \sqrt{2(\gamma_V^2-1)}\int d y {P}''(\langle X\rangle_0)}<0\,.
\end{equation}
Particularly, in the models with $P(X)\propto X^q$ and $q>1$, we find
\begin{equation}
\beta_{P,V}=-\frac{\gamma_V ^4+4 \gamma_V ^2 \left(4 q^2+q-2\right)- 12 q^2-8q+8}{\gamma ^2_V \sqrt{\gamma_V ^2-1} (2 q-1)^{5/2} (2 q+1)}<0. 
\end{equation}
Equation \eqref{eq_lift_gap_NNNLO} implies that single-phonon states are favored over multi-phonon ones.

\twocolumngrid 

\bibliographystyle{JHEP.bst}
\bibliography{Ref}

\providecommand{\href}[2]{#2}\begingroup\raggedright\begin{thebibliography}{10}

\bibitem{fetter2001vortices}
A.~L. {Fetter} and A.~A. {Svidzinsky}, \emph{{Vortices in a trapped dilute Bose-Einstein condensate}}, \href{https://doi.org/10.1088/0953-8984/13/12/201}{\emph{Journal of Physics Condensed Matter} {\bfseries 13} (2001) R135} [\href{https://arxiv.org/abs/cond-mat/0102003}{{\ttfamily cond-mat/0102003}}].

\bibitem{pethick2008bose}
C.~J. Pethick and H.~Smith, \emph{Bose--Einstein condensation in dilute gases}. Cambridge university press, 2008.

\bibitem{Schmitt:2014eka}
A.~Schmitt, \emph{{Introduction to Superfluidity}: {Field-theoretical approach and applications}}, vol.~888. 2015, \href{https://doi.org/10.1007/978-3-319-07947-9}{10.1007/978-3-319-07947-9}, [\href{https://arxiv.org/abs/1404.1284}{{\ttfamily 1404.1284}}].

\bibitem{2008LaPhy..18....1F}
A.~L. {Fetter}, \emph{{Rotating trapped Bose-Einstein condensates}}, \href{https://doi.org/10.1007/s11490-008-1001-6}{\emph{Laser Physics} {\bfseries 18} (2008) 1} [\href{https://arxiv.org/abs/0801.2952}{{\ttfamily 0801.2952}}].

\bibitem{madison2000vortex}
K.~W. {Madison}, F.~{Chevy}, W.~{Wohlleben} and J.~{Dalibard}, \emph{{Vortex Formation in a Stirred Bose-Einstein Condensate}}, \href{https://doi.org/10.1103/PhysRevLett.84.806}{\emph{\prl} {\bfseries 84} (2000) 806} [\href{https://arxiv.org/abs/cond-mat/9912015}{{\ttfamily cond-mat/9912015}}].

\bibitem{doi:10.1126/science.1060182}
J.~R. Abo-Shaeer, C.~Raman, J.~M. Vogels and W.~Ketterle, \emph{Observation of vortex lattices in bose-einstein condensates}, \href{https://doi.org/10.1126/science.1060182}{\emph{Science} {\bfseries 292} (2001) 476}.

\bibitem{tkachenko1966vortex}
V.~Tkachenko, \emph{On vortex lattices}, {\emph{Sov. Phys. JETP} {\bfseries 22} (1966) 1282}.

\bibitem{tkachenko1966stability}
V.~Tkachenko, \emph{Stability of vortex lattices}, {\emph{Sov. Phys. JETP} {\bfseries 23} (1966) 1049}.

\bibitem{tkachenko1969elasticity}
V.~Tkachenko, \emph{Elasticity of vortex lattices}, {\emph{Soviet Journal of Experimental and Theoretical Physics} {\bfseries 29} (1969) 945}.

\bibitem{sonin2014tkachenko}
E.~B. {Sonin}, \emph{{Tkachenko waves}}, \href{https://doi.org/10.1134/S0021364013240181}{\emph{Soviet Journal of Experimental and Theoretical Physics Letters} {\bfseries 98} (2014) 758} [\href{https://arxiv.org/abs/1311.1781}{{\ttfamily 1311.1781}}].

\bibitem{Moroz:2018noc}
S.~Moroz, C.~Hoyos, C.~Benzoni and D.~T. Son, \emph{{Effective field theory of a vortex lattice in a bosonic superfluid}}, \href{https://doi.org/10.21468/SciPostPhys.5.4.039}{\emph{SciPost Phys.} {\bfseries 5} (2018) 039} [\href{https://arxiv.org/abs/1803.10934}{{\ttfamily 1803.10934}}].

\bibitem{Du:2022xys}
Y.-H. Du, S.~Moroz, D.~X. Nguyen and D.~T. Son, \emph{{Noncommutative Field Theory of the Tkachenko Mode: Symmetries and Decay Rate}},  \href{https://arxiv.org/abs/2212.08671}{{\ttfamily 2212.08671}}.

\bibitem{fischer2003vortex}
U.~R. Fischer and G.~Baym, \emph{Vortex states of rapidly rotating dilute bose-einstein condensates}, {\emph{Physical review letters} {\bfseries 90} (2003) 140402}.

\bibitem{2004PhRvA..69c3608A}
A.~{Aftalion} and I.~{Danaila}, \emph{{Giant vortices in combined harmonic and quartic traps}}, \href{https://doi.org/10.1103/PhysRevA.69.033608}{\emph{\pra} {\bfseries 69} (2004) 033608} [\href{https://arxiv.org/abs/cond-mat/0309668}{{\ttfamily cond-mat/0309668}}].

\bibitem{fetter2005rapid}
A.~L. {Fetter}, B.~{Jackson} and S.~{Stringari}, \emph{{Rapid rotation of a Bose-Einstein condensate in a harmonic plus quartic trap}}, \href{https://doi.org/10.1103/PhysRevA.71.013605}{\emph{\pra} {\bfseries 71} (2005) 013605} [\href{https://arxiv.org/abs/cond-mat/0407119}{{\ttfamily cond-mat/0407119}}].

\bibitem{correggi2013vortex}
M.~{Correggi}, F.~{Pinsker}, N.~{Rougerie} and J.~{Yngvason}, \emph{{Vortex Phases of Rotating Superfluids}},  in \emph{Journal of Physics Conference Series}, vol.~414 of \emph{Journal of Physics Conference Series}, p.~012034, Feb., 2013, \href{https://arxiv.org/abs/1212.3680}{{\ttfamily 1212.3680}}, \href{https://doi.org/10.1088/1742-6596/414/1/012034}{DOI}.

\bibitem{Cuomo:2022kio}
G.~Cuomo and Z.~Komargodski, \emph{{Giant Vortices and the Regge Limit}}, \href{https://doi.org/10.1007/JHEP01(2023)006}{\emph{JHEP} {\bfseries 01} (2023) 006} [\href{https://arxiv.org/abs/2210.15694}{{\ttfamily 2210.15694}}].

\bibitem{guo2020supersonic}
Y.~{Guo}, R.~{Dubessy}, M.~d.~G. {de Herve}, A.~{Kumar}, T.~{Badr}, A.~{Perrin} et~al., \emph{{Supersonic Rotation of a Superfluid: A Long-Lived Dynamical Ring}}, \href{https://doi.org/10.1103/PhysRevLett.124.025301}{\emph{\prl} {\bfseries 124} (2020) 025301} [\href{https://arxiv.org/abs/1907.01795}{{\ttfamily 1907.01795}}].

\bibitem{gauthier2019giant}
G.~Gauthier, M.~T. Reeves, X.~Yu, A.~S. Bradley, M.~A. Baker, T.~A. Bell et~al., \emph{Giant vortex clusters in a two-dimensional quantum fluid}, {\emph{Science} {\bfseries 364} (2019) 1264}.

\bibitem{tanaka2002electronic}
K.~Tanaka, I.~Robel and B.~Jank{\'o}, \emph{Electronic structure of multiquantum giant vortex states in mesoscopic superconducting disks}, {\emph{Proceedings of the National Academy of Sciences} {\bfseries 99} (2002) 5233}.

\bibitem{dao2011giant}
V.~H. Dao, L.~Chibotaru, T.~Nishio and V.~Moshchalkov, \emph{Giant vortices, rings of vortices, and reentrant behavior in type-1.5 superconductors}, {\emph{Physical Review B---Condensed Matter and Materials Physics} {\bfseries 83} (2011) 020503}.

\bibitem{palonen2013giant}
H.~Palonen, J.~J{\"a}ykk{\"a} and P.~Paturi, \emph{Giant vortex states in type i superconductors simulated by ginzburg--landau equations}, {\emph{Journal of Physics: Condensed Matter} {\bfseries 25} (2013) 385702}.

\bibitem{PhysRevLett.59.1873}
R.~Floreanini and R.~Jackiw, \emph{Self-dual fields as charge-density solitons}, \href{https://doi.org/10.1103/PhysRevLett.59.1873}{\emph{Phys. Rev. Lett.} {\bfseries 59} (1987) 1873}.

\bibitem{elitzur1989remarks}
S.~Elitzur, G.~Moore, A.~Schwimmer and N.~Seiberg, \emph{Remarks on the canonical quantization of the chern-simons-witten theory}, {\emph{Nuclear Physics B} {\bfseries 326} (1989) 108}.

\bibitem{Jensen:2017tnb}
K.~Jensen, \emph{{Locality and anomalies in warped conformal field theory}}, \href{https://doi.org/10.1007/JHEP12(2017)111}{\emph{JHEP} {\bfseries 12} (2017) 111} [\href{https://arxiv.org/abs/1710.11626}{{\ttfamily 1710.11626}}].

\bibitem{2002PhRvB..66e4526P}
A.~{Paramekanti}, L.~{Balents} and M.~P. {Fisher}, \emph{{Ring exchange, the exciton Bose liquid, and bosonization in two dimensions}}, \href{https://doi.org/10.1103/PhysRevB.66.054526}{\emph{\prb} {\bfseries 66} (2002) 054526} [\href{https://arxiv.org/abs/cond-mat/0203171}{{\ttfamily cond-mat/0203171}}].

\bibitem{Burnell:2021reh}
F.~J. Burnell, T.~Devakul, P.~Gorantla, H.~T. Lam and S.-H. Shao, \emph{{Anomaly inflow for subsystem symmetries}}, \href{https://doi.org/10.1103/PhysRevB.106.085113}{\emph{Phys. Rev. B} {\bfseries 106} (2022) 085113} [\href{https://arxiv.org/abs/2110.09529}{{\ttfamily 2110.09529}}].

\bibitem{Son:2002zn}
D.~T. Son, \emph{{Low-energy quantum effective action for relativistic superfluids}},  \href{https://arxiv.org/abs/hep-ph/0204199}{{\ttfamily hep-ph/0204199}}.

\bibitem{Son:2005rv}
D.~T. Son and M.~Wingate, \emph{{General coordinate invariance and conformal invariance in nonrelativistic physics: Unitary Fermi gas}}, \href{https://doi.org/10.1016/j.aop.2005.11.001}{\emph{Annals Phys.} {\bfseries 321} (2006) 197} [\href{https://arxiv.org/abs/cond-mat/0509786}{{\ttfamily cond-mat/0509786}}].

\bibitem{Nicolis:2023pye}
A.~Nicolis, A.~Podo and L.~Santoni, \emph{{The connection between nonzero density and spontaneous symmetry breaking for interacting scalars}}, \href{https://doi.org/10.1007/JHEP09(2023)200}{\emph{JHEP} {\bfseries 09} (2023) 200} [\href{https://arxiv.org/abs/2305.08896}{{\ttfamily 2305.08896}}].

\bibitem{Nishida:2007pj}
Y.~Nishida and D.~T. Son, \emph{{Nonrelativistic conformal field theories}}, \href{https://doi.org/10.1103/PhysRevD.76.086004}{\emph{Phys. Rev. D} {\bfseries 76} (2007) 086004} [\href{https://arxiv.org/abs/0706.3746}{{\ttfamily 0706.3746}}].

\bibitem{2006PhRvA..74e3604W}
F.~{Werner} and Y.~{Castin}, \emph{{Unitary gas in an isotropic harmonic trap: Symmetry properties and applications}}, \href{https://doi.org/10.1103/PhysRevA.74.053604}{\emph{\pra} {\bfseries 74} (2006) 053604} [\href{https://arxiv.org/abs/cond-mat/0607821}{{\ttfamily cond-mat/0607821}}].

\bibitem{Nishida:2010tm}
Y.~Nishida and D.~T. Son, \emph{{Unitary Fermi gas, epsilon expansion, and nonrelativistic conformal field theories}}, \href{https://doi.org/10.1007/978-3-642-21978-8_7}{\emph{Lect. Notes Phys.} {\bfseries 836} (2012) 233} [\href{https://arxiv.org/abs/1004.3597}{{\ttfamily 1004.3597}}].

\bibitem{Hellerman:2015nra}
S.~Hellerman, D.~Orlando, S.~Reffert and M.~Watanabe, \emph{{On the CFT Operator Spectrum at Large Global Charge}}, \href{https://doi.org/10.1007/JHEP12(2015)071}{\emph{JHEP} {\bfseries 12} (2015) 071} [\href{https://arxiv.org/abs/1505.01537}{{\ttfamily 1505.01537}}].

\bibitem{Monin:2016jmo}
A.~Monin, D.~Pirtskhalava, R.~Rattazzi and F.~K. Seibold, \emph{{Semiclassics, Goldstone Bosons and CFT data}}, \href{https://doi.org/10.1007/JHEP06(2017)011}{\emph{JHEP} {\bfseries 06} (2017) 011} [\href{https://arxiv.org/abs/1611.02912}{{\ttfamily 1611.02912}}].

\bibitem{Kravec:2018qnu}
S.~M. Kravec and S.~Pal, \emph{{Nonrelativistic Conformal Field Theories in the Large Charge Sector}}, \href{https://doi.org/10.1007/JHEP02(2019)008}{\emph{JHEP} {\bfseries 02} (2019) 008} [\href{https://arxiv.org/abs/1809.08188}{{\ttfamily 1809.08188}}].

\bibitem{Kravec:2019djc}
S.~M. Kravec and S.~Pal, \emph{{The Spinful Large Charge Sector of Non-Relativistic CFTs: From Phonons to Vortex Crystals}}, \href{https://doi.org/10.1007/JHEP05(2019)194}{\emph{JHEP} {\bfseries 05} (2019) 194} [\href{https://arxiv.org/abs/1904.05462}{{\ttfamily 1904.05462}}].

\bibitem{Hellerman:2021qzz}
S.~Hellerman, D.~Orlando, V.~Pellizzani, S.~Reffert and I.~Swanson, \emph{{Nonrelativistic CFTs at large charge: Casimir energy and logarithmic enhancements}}, \href{https://doi.org/10.1007/JHEP05(2022)135}{\emph{JHEP} {\bfseries 05} (2022) 135} [\href{https://arxiv.org/abs/2111.12094}{{\ttfamily 2111.12094}}].

\bibitem{Hellerman:2020eff}
S.~Hellerman and I.~Swanson, \emph{{Droplet-Edge Operators in Nonrelativistic Conformal Field Theories}},  \href{https://arxiv.org/abs/2010.07967}{{\ttfamily 2010.07967}}.

\bibitem{Cuomo:2021cnb}
G.~Cuomo, M.~Mezei and A.~Raviv-Moshe, \emph{{Boundary conformal field theory at large charge}}, \href{https://doi.org/10.1007/JHEP10(2021)143}{\emph{JHEP} {\bfseries 10} (2021) 143} [\href{https://arxiv.org/abs/2108.06579}{{\ttfamily 2108.06579}}].

\bibitem{Horn:2015zna}
B.~Horn, A.~Nicolis and R.~Penco, \emph{{Effective string theory for vortex lines in fluids and superfluids}}, \href{https://doi.org/10.1007/JHEP10(2015)153}{\emph{JHEP} {\bfseries 10} (2015) 153} [\href{https://arxiv.org/abs/1507.05635}{{\ttfamily 1507.05635}}].

\bibitem{Kourkoulou:2023xqe}
I.~Kourkoulou, M.~J. Landry, A.~Nicolis and K.~Parmentier, \emph{{Apparently superluminal superfluids}}, \href{https://doi.org/10.1007/JHEP01(2024)080}{\emph{JHEP} {\bfseries 01} (2024) 080} [\href{https://arxiv.org/abs/2305.11226}{{\ttfamily 2305.11226}}].

\bibitem{Evans:2020uui}
G.~W. Evans and A.~Schmitt, \emph{{Strange quark mass turns magnetic domain walls into multi-winding flux tubes}}, \href{https://doi.org/10.1088/1361-6471/abcb9d}{\emph{J. Phys. G} {\bfseries 48} (2021) 035002} [\href{https://arxiv.org/abs/2009.01141}{{\ttfamily 2009.01141}}].

\bibitem{Penin:2020cxj}
A.~A. Penin and Q.~Weller, \emph{{What Becomes of Giant Vortices in the Abelian Higgs Model}}, \href{https://doi.org/10.1103/PhysRevLett.125.251601}{\emph{Phys. Rev. Lett.} {\bfseries 125} (2020) 251601} [\href{https://arxiv.org/abs/2009.06640}{{\ttfamily 2009.06640}}].

\bibitem{Gates:2022bnv}
L.~Gates and A.~A. Penin, \emph{{Majorana modes of giant vortices}}, \href{https://doi.org/10.1103/PhysRevB.107.125418}{\emph{Phys. Rev. B} {\bfseries 107} (2023) 125418} [\href{https://arxiv.org/abs/2210.04908}{{\ttfamily 2210.04908}}].

\bibitem{PhysRev.153.285}
A.~L. Fetter, \emph{Low-lying superfluid states in a rotating annulus}, \href{https://doi.org/10.1103/PhysRev.153.285}{\emph{Phys. Rev.} {\bfseries 153} (1967) 285}.

\bibitem{2023PhRvA.107e3317R}
A.~{Richaud}, G.~{Lamporesi}, M.~{Capone} and A.~{Recati}, \emph{{Mass-driven vortex collisions in flat superfluids}}, \href{https://doi.org/10.1103/PhysRevA.107.053317}{\emph{\pra} {\bfseries 107} (2023) 053317} [\href{https://arxiv.org/abs/2209.00493}{{\ttfamily 2209.00493}}].

\bibitem{future}
G.~Cuomo, Z.~Komargodski, A.~Nicolis and S.~Zhong, \emph{In progress}, .

\bibitem{Gibbons:2010fb}
G.~W. Gibbons and C.~N. Pope, \emph{{Kohn's Theorem, Larmor's Equivalence Principle and the Newton-Hooke Group}}, \href{https://doi.org/10.1016/j.aop.2011.03.003}{\emph{Annals Phys.} {\bfseries 326} (2011) 1760} [\href{https://arxiv.org/abs/1010.2455}{{\ttfamily 1010.2455}}].

\bibitem{vallis_2006}
G.~K. Vallis, \emph{Atmospheric and Oceanic Fluid Dynamics: Fundamentals and Large-scale Circulation}. Cambridge University Press, 2006, \href{https://doi.org/10.1017/CBO9780511790447}{10.1017/CBO9780511790447}.

\bibitem{Tong:2022gpg}
D.~Tong, \emph{{A gauge theory for shallow water}}, \href{https://doi.org/10.21468/SciPostPhys.14.5.102}{\emph{SciPost Phys.} {\bfseries 14} (2023) 102} [\href{https://arxiv.org/abs/2209.10574}{{\ttfamily 2209.10574}}].

\bibitem{2023arXiv230305669M}
G.~M. {Monteiro} and S.~{Ganeshan}, \emph{{Coastal Kelvin Mode and the Fractional Quantum Hall Edge}}, \href{https://doi.org/10.48550/arXiv.2303.05669}{\emph{arXiv e-prints} (2023) arXiv:2303.05669} [\href{https://arxiv.org/abs/2303.05669}{{\ttfamily 2303.05669}}].

\bibitem{sonin2016dynamics}
E.~Sonin, \emph{Dynamics of Quantised Vortices in Superfluids}. Cambridge University Press, 2016.

\bibitem{2005PhRvL..94j0402C}
M.~{Cozzini}, A.~L. {Fetter}, B.~{Jackson} and S.~{Stringari}, \emph{{Oscillations of a Bose-Einstein Condensate Rotating in a Harmonic Plus Quartic Trap}}, \href{https://doi.org/10.1103/PhysRevLett.94.100402}{\emph{\prl} {\bfseries 94} (2005) 100402} [\href{https://arxiv.org/abs/cond-mat/0411106}{{\ttfamily cond-mat/0411106}}].

\bibitem{Guica:2008mu}
M.~Guica, T.~Hartman, W.~Song and A.~Strominger, \emph{{The Kerr/CFT Correspondence}}, \href{https://doi.org/10.1103/PhysRevD.80.124008}{\emph{Phys. Rev. D} {\bfseries 80} (2009) 124008} [\href{https://arxiv.org/abs/0809.4266}{{\ttfamily 0809.4266}}].

\bibitem{Hofman:2011zj}
D.~M. Hofman and A.~Strominger, \emph{{Chiral Scale and Conformal Invariance in 2D Quantum Field Theory}}, \href{https://doi.org/10.1103/PhysRevLett.107.161601}{\emph{Phys. Rev. Lett.} {\bfseries 107} (2011) 161601} [\href{https://arxiv.org/abs/1107.2917}{{\ttfamily 1107.2917}}].

\bibitem{Nicolis:2012vf}
A.~Nicolis and F.~Piazza, \emph{{Implications of Relativity on Nonrelativistic Goldstone Theorems: Gapped Excitations at Finite Charge Density}}, \href{https://doi.org/10.1103/PhysRevLett.110.011602}{\emph{Phys. Rev. Lett.} {\bfseries 110} (2013) 011602} [\href{https://arxiv.org/abs/1204.1570}{{\ttfamily 1204.1570}}].

\bibitem{Watanabe:2013uya}
H.~Watanabe, T.~Brauner and H.~Murayama, \emph{{Massive Nambu-Goldstone Bosons}}, \href{https://doi.org/10.1103/PhysRevLett.111.021601}{\emph{Phys. Rev. Lett.} {\bfseries 111} (2013) 021601} [\href{https://arxiv.org/abs/1303.1527}{{\ttfamily 1303.1527}}].

\end{thebibliography}\endgroup

\end{document}